\documentclass[useAMS,usenatbib]{STYFiles/mn2e}
\usepackage{graphicx,amsmath}
\newcommand{\HI}{H\,{\sc i}}

\date{Released 2014 Xxxxx XX}
\pubyear{2014}

\begin{document}

\title[ALFALFA Void HIMF \& WF]{The \HI\ Mass Function and Velocity Width Function of Void Galaxies in the Arecibo Legacy
  Fast ALFA Survey}

\author[Moorman et al.]{
Crystal M. Moorman,$^1$ Michael S. Vogeley,$^1$ 
Fiona Hoyle,$^2$ Danny C. Pan,$^3$
 \newauthor
Martha P. Haynes,$^4$ Riccardo Giovanelli$^4$ \\
1. Department of Physics, Drexel University,
3141 Chestnut Street, Philadelphia, PA 19104, USA \\ 
2. Pontifica Universidad Catolica de Ecuador, 12 de Octubre 1076 y Roca, Quito, Ecuador \\
3. Shanghai Astronomical Observatory, Shanghai, China, 200030 \\
4. Center for Radiophysics and Space Research, Space Sciences Building, Cornell University Ithaca, NY 14853 \\
}

\maketitle 
\begin{abstract}
We measure the \HI\ mass function (HIMF) and velocity width function (WF) 
across environments over a range of masses,
$7.2<\log\left(M_{\rm HI}/M_{\odot}\right)<10.8$, and profile widths, $1.3\log($km s$^{-1})<\log\left(W\right)<2.9\log($km s$^{-1})$,
using a catalog of $\sim$7,300 \HI-selected galaxies from the ALFALFA Survey,
located in the region of sky where ALFALFA and SDSS (Data Release 7) North overlap. 
We divide our galaxy sample into those that reside in large-scale voids (void galaxies) and 
those that live in denser regions (wall galaxies). 
We find the void HIMF to be well fit by a Schechter function with normalization 
$\Phi^*=(1.37\pm0.1) \times 10^{-2} h^3$Mpc$^{-3}$, 
characteristic mass log($M_{\rm HI}^*/M_{\odot}$)+ 2log$h_{70}=9.86\pm0.02$, 
and low-mass-end slope $\alpha=-1.29\pm0.02$.
Similarly, for wall galaxies, we find best-fitting parameters $\Phi^*=(1.82\pm0.03) \times 10^{-2} h^3$Mpc$^{-3}$, 
log($M_{\rm HI}^*/M_{\odot}$)+ 2log$h_{70}=10.00\pm0.01$, and $\alpha=-1.35\pm0.01$.
We conclude that void galaxies typically have slightly lower \HI\ masses than 
their non-void counterparts,  
which is in agreement with the dark matter halo mass function shift in voids 
assuming a simple relationship between DM mass and \HI\ mass. 
We also find that the low-mass slope of the void HIMF is similar to that of the wall HIMF 
suggesting that there is either no excess of low-mass galaxies in voids or there is an 
abundance of intermediate \HI\ mass galaxies. 
We fit a modified Schechter function to the ALFALFA void WF and determine its best-fitting parameters 
to be $\Phi^*=0.21\pm0.1 h^3$Mpc$^{-3}$, 
$\log(W^*)=2.13\pm0.3$, $\alpha=0.52\pm0.5$ and high-width slope $\beta=1.3\pm0.4$.  
For wall galaxies, the WF parameters are: $\Phi^*=0.022\pm0.009 h^3$Mpc$^{-3}$, 
$\log(W^*)=2.62\pm0.5$, $\alpha=-0.64\pm0.2$ and $\beta=3.58\pm1.5$.
Because of large uncertainties on the void and wall width functions, we cannot conclude whether 
the WF is dependent on the environment. 
\end{abstract}


\section{Introduction}
The advent of large redshift surveys
has allowed for a detailed mapping of the nearby Universe, revealing 
a variety of environments within which galaxies reside: 
rich clusters, groups, filaments, and the underdense regions filling the volume between. 
The starkly underdense regions filling most of our Universe, called ``voids'',
are relatively pristine environments for studying galaxy evolution and formation, 
because gas-stripping galaxy interactions are exceptionally rare. 
The study of galaxies living in these dynamically-distinct environments is therefore 
crucial to understanding the processes that affect galaxy formation.  

A variety of definitions of ``voids" exists in the literature 
\citep{ElAd1997,ElAdPiran1997,ElAdPiran2000,Plionis2002,Sheth2004,
HoyleVogeley2004,Rojas2004,Blanton2005a,Patiri2006,
vonBenda-Beckmann2007,Melnyk2009,Karachentseva2010,
Park2012,Sutter2012,Elyiv2013} 
ranging anywhere from locally isolated galaxies 
to extreme large-scale underdensities. We wish to focus on large-scale structures 
and define voids as regions with density contrast $\delta<-$0.8 in optically-selected 
bright galaxies ($\sim L^*$) with a minimum radius of 10$h^{-1}$Mpc.
That is, voids are not samples of ``isolated'' galaxies identified by visual inspection 
(e.g. \citealt{Melnyk2009, Karachentseva2010}).
Large voids with density contrast $\delta<-0.8$ naturally arise via gravitational instability \citep{Sheth2004}.
These large-scale voids may be identified using a variety of void-finding algorithms 
including 
\cite{Kauffmann1991,ElAdPiran1997,Aikio1998,SchaapvandeWeygaert2000,HoyleVogeley2002,Neyrinck2008,AragonCalvo2010}. 
A comparison of different void-finding techniques may be found in \citealt{Colberg2008}. 
Studies reveal that large-scale voids occupy approximately 60 per cent of 
the volume of our Universe (\citealt{Pan2012})
and the galaxies within these voids form substructure that is evident 
in both simulations and observations \citep{MathisWhite2002,Benson2003,Gottlober2003,
vandeWeygaert2010Depressed,AragonCalvo2013}.

Simulations of the $\Lambda$CDM cosmological model 
predict an abundance of low-mass dark matter haloes (DMH) across
all environments with a shift in the DMH mass function to lower masses in voids \citep{Goldberg2005}.
If we assume a linear relationship between DMH mass, galaxy light 
and baryonic mass, then we expect to see an abundance of 
low-mass, low-luminosity galaxies in voids and denser regions as well. 
We know this assumption to be false, because 
star formation is suppressed at both high- and low-mass ends by e.g. AGN feedback and supernovae feedback, respectively. 
Given the phenomena affecting the low-mass ends, one might expect 
the low-mass slopes to vary across environment. 
For instance, ram pressure stripping removes cold gas from low-mass
galaxies in clusters but this phenomenon would not be as prevalent in void galaxies. 
One might also expect the effects from supernovae to set in at a particular mass, yet 
it seems to affect both the void and wall galaxies similarly, regardless of the shift in 
characteristic mass between environment. 
To date, there is no evidence for a difference between the low-mass slopes in voids vs. denser regions,
but we find this odd given the phenomena affecting the low-mass slopes. 
Critical points of comparison between the $\Lambda$CDM structure formation model 
and galaxy observations are found in the faint end of the 
galaxy luminosity function (LF) and 
the low-mass end of the neutral hydrogen mass function (HIMF).
The global DMH mass function is predicted to follow a Schechter function 
\citep{PressSchechter1974} with a very steep low-mass end slope ($\alpha\simeq-1.8$;
\cite{MathisWhite2002}), but 
measurements of the observed faint and low-mass end slopes of the global galaxy LF 
(e.g. \cite{Blanton2005ELL}) and HIMF (e.g. \cite{Martin2010}) 
are significantly shallower ($\alpha\simeq-1.3$). 
While \cite{Blanton2005ELL} note that selection effects of low-surface brightness galaxies 
could cause the slope to flatten, their estimate of the corrected slope is still too shallow ($\alpha\simeq-1.5$) 
to match DMH predictions. 

Studies focusing on galaxies in large-scale voids 
indicate that void galaxies are generally bluer, fainter, later-type, and have higher 
star formation rates (SFRs) per stellar mass than their counterparts 
in denser regions \citep{Rojas2004, Rojas2005,Hoyle2005,Blanton2005a,
Croton2005,Park2007,CrotonFarrar2008,Cen2011,Kreckel2011,Hoyle2012, Geha2012}. 
Optical observations also reveal a discrepancy between 
the number of low-luminosity/low-mass (dwarf) 
galaxies and the predicted number of low-mass haloes (e.g. \citealt{Karachentsev2004}).
This discrepancy is part of the void problem mentioned in \cite{Peebles2001} which 
states that all types of galaxies appear to respect the voids, whereas 
$\Lambda$CDM predicts that voids should contain an abundance of low mass objects.
$\Lambda$CDM simulations \citep{Warren2006,Hoeft2006} predict 
that voids have a density of low-mass haloes that is 1/10 that of the cosmic mean. 
This is consistent with bright galaxies in large scale voids \citep{HoyleVogeley2004,
Conroy2005, Hoeft2006,Tinker2008}; 
however, optical observations do not reveal a plethora of faint galaxies in voids 
(e.g. \cite{Karachentsev2004}; in the Local Void). 
Among the attempts to identify an abundance of dwarf galaxies in voids
are \cite{GroginGeller99,Blanton2003LF,Hoyle2005,Croton2005} and \cite{Hoyle2012} who 
investigate the environmental dependence of the 
LF of optically-selected galaxies. These results vary depending on 
the definition of environment used in the work, and none identifies an excess of 
dwarf galaxies in extreme voids. 
\cite{Koposov2009} propose that the lack of dwarf galaxies in voids found in optically-selected 
samples could be due to strong star formation 
suppression of void galaxies, before and after reionization, as well as 
observational selection effects. If these effects are responsible, we would expect 
a population of low-mass, optically-dark, yet \HI-rich galaxies to exist. 
Such a population should be detectable by their gas. 

The results of \cite{Basilakos2007} indicate that low \HI-mass galaxies
seem to avoid underdense regions. 
To date, with only a handful of exceptions (e.g. \citealt{Giovanelli2013}), evidence of a population 
of gas-rich, optically dark galaxies large enough to reconcile dwarf galaxy counts 
with $\Lambda$CDM models does not exist \citep{Haynes2008}. 
The first generation of blind \HI\ surveys were typically shallow and 
unable to detect \HI\ clouds over a large range of masses. 
Small number statistics and uncertainties in distances of nearby 
galaxies \citep{Masters2004} made determining large scale environmental effects 
on galactic \HI\ content difficult. 
\cite{Haynes1984} show that environment has an impact on the \HI\ content 
of galaxies, where galaxies in clusters tend to be more \HI\ deficient than ``field'' galaxies. 
Using the Arecibo Dual Beam Survey \citep{RosenbergSchneider2000},
\cite{RosenbergSchneider2002} corroborate this work hinting 
at the influence of large scale structure on the HIMF. These authors show that, 
for galaxies in the Virgo Cluster, the low-mass-end slope of the HIMF becomes shallower.

With the onset of deeper, large area blind-\HI\ surveys, we are able to better constrain the environmental 
impact on a galaxy's \HI\ content. 
Headway has been made on determining the environmental dependence 
of the HIMF of \HI\ clouds from second generation surveys by \cite{Zwaan2005} using the 
\HI\ Parkes All Sky Survey (HIPASS; \citealt{Meyer2004}) and 
by \cite{Springob2005}, \cite{Stierwalt2009}, and \cite{Martin2011} using various data releases 
from the ALFALFA Survey \citep{ALFALFAII,ALFALFAI}. 
We find in the literature that the low-mass end slope of the HIMF 
may become steeper or shallower with density depending on sample selection and the 
definition of environment. See, for example, \cite{Zwaan2005}, \cite{Springob2005}, 
\cite{Toribio2011LowDensity}, and \cite{Martin2011} 
who use either \HI- or optically-selected samples to define the environment of \HI-selected galaxies  
on scales ranging from locally isolated galaxies to low density environments on scales of $\sim$10Mpc. 
We discuss the conflicting HIMF results in more detail below in Section \ref{subsec:himf_comp}. 

Another point of comparison between $\Lambda$CDM simulations and observations 
is through estimating corrections to the velocity width function (WF) to obtain a circular velocity 
function, although this comparison is less direct. 
Cold dark matter models predict the circular 
velocity functions of haloes to follow a power law with a steep slope of 
$\alpha\sim-3$ \citep{Klypin1999, Zavala2009}.
Observations do not confirm this prediction. In fact, 
the observed velocity functions closely resemble Schechter functions 
with much shallower low-velocity slopes 
(e.g. \citealt{Sheth2003,Zwaan2010,Papastergis2011}). 
The observed velocity function of galaxies may be obtained indirectly 
through optical photometry or spectra 
using of the Tully-Fisher relation 
\citep{Desai2004, Abramson2013} or through 
direct observations of the velocity width using \HI\ surveys
(\cite{Zwaan2010} with HIPASS; \cite{Papastergis2011} with ALFALFA). 

A comparison of previous work on both optically-selected void galaxies and \HI-selected galaxies reveal 
similar characteristics between these two samples. 
\cite{Toribio2009} investigate 
properties of ``isolated'' \HI-selected galaxies from ALFALFA and find 
that these remote objects tend to be blue, late-type, star-forming galaxies. 
\cite{Huang2012} compare properties of \HI- and optically-selected 
samples found within the volume covered by both SDSS DR7 and ALFALFA 
and find that for a given stellar mass, \HI-selected galaxies generally 
have higher star formation rates and specific star formation rates, yet lower star formation efficiencies. 
\cite{Rojas2004,Hoyle2005,Hoyle2012} find that optically-selected void galaxies are generally bluer, fainter, late-type, and have 
higher specific star formation rates than galaxies in denser regions.
Given the similarities in the characteristics of galaxies detected by blind \HI\ surveys and void 
galaxies found using optically-selected data, we would expect that 
\HI\ surveys yield a higher fraction of void galaxy detections than that of galaxies found using optical surveys.
As discussed in more detail below, we find that 
33 per cent of ALFALFA detections reside in large scale voids, while 
in the same volume (out to $z \sim 0.5$) 26 per cent of magnitude-limited SDSS detections are classified as void galaxies.

In this paper, we study 
the HIMF and WF of ALFALFA galaxies that 
lie in deep large-scale voids. 
In Section \ref{sec:data} we discuss 
our void catalog and \HI-selected sample. 
In Section \ref{sec:meth} we present our 
methods and results for the HIMF and WF of void galaxies. 
We compare our results 
to previous work in Section \ref{sec:discuss} and summarize our results in 
Section \ref{sec:conc}. 
Unless otherwise specified, we adopt the distances, $D_i$; \HI\ masses, $M_{\rm HI}$; 
velocity widths, $W_{50}$; and integrated \HI\ fluxes, $S_{int}$ reported 
in the $\alpha$.40 catalog \citep{Haynes2011}, as well as the ALFALFA adopted 
Hubble constant $h=H_0$/100 km s$^{-1}$Mpc$^{-1}=0.7$. 
Where comoving coordinates are determined, we assume 
$\Omega_m=0.26$ and $\Omega_{\Lambda}=0.74$.

\section{Data}
\label{sec:data}
\subsection{Finding A Void Galaxy Sample Using SDSS DR7}
\label{subsec:SDSSdata}
The Sloan Digital Sky Survey Data Release 7 (SDSS DR7) \citep{Abazajian2009} is a wide-field
multi-band imaging and spectroscopic survey that uses drift scanning
to map 8,032 deg$^2$ of the northern sky. SDSS utilizes the
2.5m telescope located at Apache Point Observatory in New Mexico,
allowing it to cover $\sim10^4$ deg$^2$ of the northern hemisphere in the five band SDSS
system-$u,g,r,i,$ and $z$ \citep{Fukugita1996, Gunn1998}.  
Galaxies with Petrosian $r$-band magnitude $r < 17.77$ are selected for 
spectroscopic follow up \citep{Lupton2001,Strauss2002}. 
Spectra obtained through the SDSS are taken
using two double fiber-fed spectrographs and fiber plug plates covering
a portion of the sky $1.49^{\circ}$ in radius with a minimum fiber 
separation of 55 arcseconds \citep{Blanton2003SpectraSDSS}.

The Korea Institute for Advanced Study Value-Added Galaxy
Catalog (KIAS-VAGC) of \cite{Choi2010} is based on the 
SDSS DR7. This consists of 583,946 galaxies 
with $10 < r < 17.6$ from the 
NYU-VAGC Large Scale Structure Sample (brvoid0) \citep{Blanton2005NYUVAGC}, 
114,303 galaxies with $17.6 < r < 17.77$ from NYU-VAGC
(full0), and 10,497 galaxies from either UZC, PSCz, RC3, or 2dF which
were excluded by SDSS. We omit 929 objects, mostly de-blended outlying parts 
of large galaxies,  for a total of 707,817 galaxies. 

To create a void catalog, \cite{Pan2012} employ the void finding algorithm of 
\cite{HoyleVogeley2002}, called VoidFinder -- based on the \cite{ElAdPiran1997}
approach -- on a volume limited sample of the 
KIAS-VAGC. 
The volume-limited sample we use consists of 120,606 galaxies 
within $z=0.107$ corresponding to an absolute magnitude limit of $M_r<-20.09$.
VoidFinder uses a nearest neighbors algorithm to identify volume-limited 
galaxies in low density regions. If a galaxy's third nearest neighbor is at 
least 6.3$h^{-1}$ Mpc away it is considered a potential void galaxy and is 
removed from the sample. VoidFinder then grows spheres in the empty 
spaces until the spheres are bounded by four remaining galaxies. If the 
sphere has a radius larger than 10$h^{-1}$ Mpc it is considered a void, otherwise it is discarded. 
Each void is comprised of multiple spheres and we define the centre of each void to be the 
centre of the sphere with the maximal radius.
The VoidFinder parameters were chosen to select void 
regions with density contrast $\delta<-$0.8.
For a more detailed description of how VoidFinder works, 
see \cite{HoyleVogeley2002} and \cite{Pan2012}.

\begin{figure}
  \begin{center}
    \includegraphics[scale=0.19]{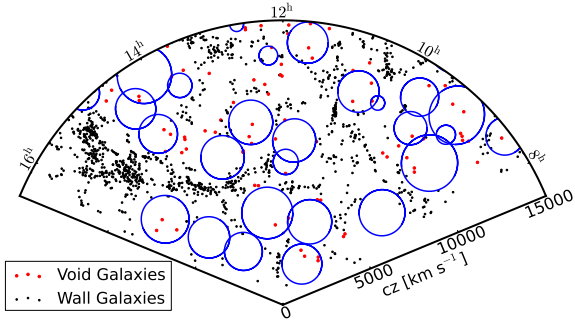}
  \end{center}
  \caption{\small 10$h^{-1}$Mpc slab of SDSS DR7 volume-limited sample.
  Void galaxies are displayed as red points, and wall galaxies are displayed 
   as black points. The circles depict the intersection of only the maximal sphere of a 
   void with the centre of the slice. Note that voids tend to be ellipsoidal rather than 
   spherical, thus void galaxies may appear outside of the maximal sphere drawn.
    \normalsize}
  \label{fig:vollimslice}
\end{figure}

\cite{Pan2012} identify 1,054 voids with radii greater than 10$h^{-1}$ Mpc 
occupying approximately 60 per cent of the volume covered by the SDSS DR7 
out to z=0.107. 
These voids have less than 10 per cent of the average density out to the walls. 
They are ellipsoidal in shape with a preference for being prolate.
The median effective void radius is 15$h^{-1}$ Mpc with over half of the volume 
consisting of voids with effective radius greater than 17.8$h^{-1}$ Mpc.
Figure \ref{fig:vollimslice} shows a 10$h^{-1}$ Mpc thick redshift slice 
of the volume-limited sample used to identify large-scale voids, 
centred at R.A.=12$^h$, Dec=10$^{\circ}$. Wall galaxies in the volume-limited sample 
are shown as black points and void galaxies are shown as red points. 
The circles depict the intersection of the maximal spheres of each void with the
centre of the slice.

\begin{figure}
  \begin{center}
    \includegraphics[scale=0.19]{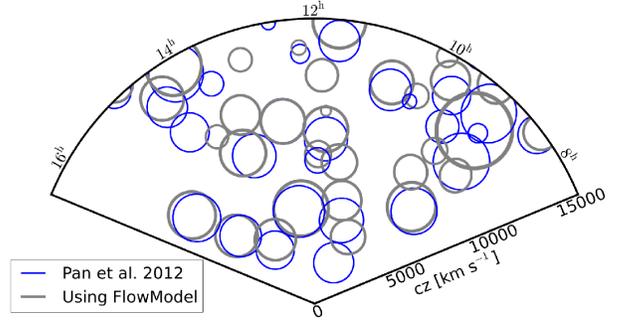}
  \end{center}
  \caption{\small We process the KIAS data set through 
  the FlowModel of Masters et al. (2005) and obtain a new volume-limited sample to run through 
  VoidFinder.
  This figure depicts the intersection of the centre of a 10$h^{-1}$Mpc slab 
  centred at R.A.=12$^h$, Dec=10$^{\circ}$ 
  with the resulting maximal spheres of this new set of voids (gray circles). For comparison, 
  we also show the intersection of the slab with the maximal void spheres of the 
  Pan et al. 2012 void catalog (blue circles) found using an SDSS DR7 volume-limited sample.
    \normalsize}
  \label{fig:compareflowmodel}
\end{figure}

\begin{figure}
  \begin{center}
    \includegraphics[scale=0.2]{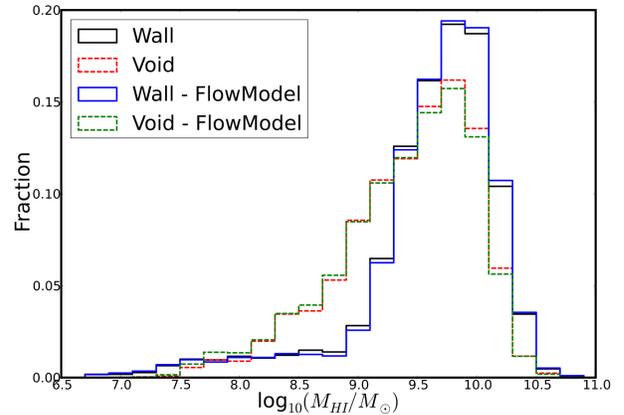} 
  \end{center}
  \caption{\small 
  \HI\ mass distribution of void (red dashed line) and wall (black solid line) galaxies as 
  determined by the Pan et al. (2012) 
  void catalog and the \HI\ mass distribution of void (green dashed line) and wall (blue solid line) 
  galaxies as determined by the modified void catalog obtained 
  using the FlowModel of Masters et al. (2005).
    \normalsize}
  \label{fig:void_compareflow}
\end{figure}

To account for the effects of the local peculiar velocity field, 
we apply the FlowModel of Masters (2005) to 
the KIAS-VAGC galaxies. We then create a volume-limited sample with 
absolute magnitude limit $M_r<-20.09$. We apply the VoidFinder algorithm 
to this volume-limited sample. See Figure \ref{fig:compareflowmodel} for the resulting void locations 
after applying the FlowModel to the KIAS sample compared to those of \cite{Pan2012}. 
This figure depicts the same region of sky shown in Figure \ref{fig:vollimslice}
with the intersection of the centre of the slab with the modified void catalog alongside the 
intersection of the slab with the \cite{Pan2012} void catalog. 
Comparing galaxy locations to the void catalog found using the FlowModel does not result in a 
significant difference in void and wall galaxy samples (see Figure \ref{fig:void_compareflow}).
Because we do not notice a significant difference between 
the void catalog of \cite{Pan2012} and our void catalog, we use the void catalog 
of \cite{Pan2012} for consistency. 

\subsection{The ALFALFA Sample}
\label{subsec:ALFALFAdata}

The Arecibo Legacy Fast ALFA (ALFALFA) Survey is a large-area, blind 
extragalactic \HI\ survey that will detect $>$30,000 galaxies out to $cz\sim18000$km s$^{-1}$ 
with a median redshift of $\sim$8,000 km s$^{-1}$, over 7000 deg$^2$ of sky upon completion.
ALFALFA has a 5$\sigma$ detection limit of 0.72 Jy km s$^{-1}$ for a source with a profile 
width of 200 km s$^{-1}$ \citep{ALFALFAII} and allows for the detection of galaxies
with \HI\ masses down to $M_{\rm HI}=10^8 M_{\odot}$ out to 40 Mpc. 
The most recent release of the ALFALFA Survey \citep{ALFALFAII,ALFALFAI}, 
the $\alpha.40$ catalog of 
\cite{Haynes2011}, covers $\sim$2800 deg$^2$, 
approximately 40 per cent of the final survey area. This catalog,
which consists of 15,041 \HI\ detections and contains previously released catalogs 
\citep{Giovanelli2007, Saintonge2008, Kent2008,
Stierwalt2009, Martin2009},
covers two regions in the northern Galactic
Hemisphere, referred to as the Spring Sky, 
($07^h30^m<$R.A.$<16^h30^m$, $04^{\circ}<$ Dec
$<16^{\circ}$ and  $24^{\circ}<$ Dec $<28^{\circ}$), and two in the 
southern Galactic Hemisphere, referred to as the Fall Sky, 
($22^h<$R.A.$<03^h$, $14^{\circ}<$ Dec
$<16^{\circ}$ and  $24^{\circ}<$ Dec $<32^{\circ}$). 
\cite{Haynes2011} categorize the confidently detected \HI\ sources of 
the $\alpha.40$ catalog into one of three categories:
Code 1 objects are reliable detections with high signal-to-noise ratio (S/N$>$6.5),
Code 2 objects have lower S/N, but coincide with optical counterparts with known 
redshift similar to that of the \HI\ detection, and Code 9 objects which correspond to 
high velocity clouds (HVCs). 

Our interest lies in identifying reliable \HI\ detections in voids. To identify $R>10h^{-1}$Mpc voids, 
VoidFinder requires wide angular coverage in spectroscopy, which for SDSS DR7, is 
only possible in North. This region of sky corresponds to the ALFALFA Spring Sky, thus 
we reduce our sample to objects included only in the Spring Sky. 
We further limit ourselves to Code 1 detections 
which lie within the redshift range cz $\le15000$km s$^{-1}$; beyond this redshift range, 
the FAA radar at the San Juan Airport interferes with ALFALFA's
detection ability over a range of frequencies corresponding to a shell
of thickness $\sim$10 Mpc. Our final sample contains 8,118 \HI\ sources over an area 
of 2,077 deg$^2$ corresponding to $\sim30$ per cent of 
the final projected survey area.

To ensure that we are using an unbiased \HI-selected sample, we use the aforementioned 
subset of the full $\alpha.40$ data set to determine the HIMF and WF of void galaxies. 
However, certain tasks we would like to accomplish (namely those in Sections \ref{subsubsec:WF}) require 
that we cross-match the \HI-detections with optical galaxies.
\cite{Haynes2011} supply a 
cross-reference of 12,468 ALFALFA \HI\ sources with the most probable 
optical counterpart from the SDSS DR7 where the two survey footprints overlap. 
Of the original 15,041 \HI\ detections in the $\alpha.40$ catalog, 201 have no assigned 
optical counterpart, 2312 lie outside the SDSS DR7 footprint, and 60 appear within 
the SDSS DR7 imaging survey but the galaxies' images are contaminated by bright foreground stars or other artefacts. 
From the cross-referenced catalog, we obtain photometric object IDs and 
query the SDSS DR7 database to obtain information regarding the apparent magnitude in each Sloan filter 
for every object in order to identify the objects' colour and absolute $r$-band magnitude, $M_r$. 
Because we are comparing optically-selected galaxy LFs to \HI-selected galaxy LFs, we wish to remain 
consistent in how we determine absolute magnitudes; therefore, we $K$-correct the magnitudes and 
band-shift each \HI\ source's $M_r$ to $z=0.1$ using $K$-correct Version 4.4 \citep{Blanton_kcorrect} 
as done in the KIAS-VAGC.

\section{Methods} 
\label{sec:meth}

\subsection{Creating a Void \HI-Selected Sample}
\label{subsec:creating_void_sample}
We categorize the \HI\ sources within our sample into void and wall galaxy samples
by comparing the coordinates of each galaxy to 
the void catalog of \cite{Pan2012}. 
The void locations are found using comoving coordinates; to ensure we are 
consistent in finding the locations of \HI\ clouds with respect to the voids, we 
use the redshift of each detection to obtain its comoving coordinates in 
$h^{-1}$Mpc.
From our \HI-detected sample, we identify $2,777$ (33\%) void galaxies and $4,857$ (60\%) wall galaxies.
The remaining 384 (6\%) ALFALFA detections lie near the 
edges of the SDSS DR7 mask, so we cannot determine whether the galaxies 
live in a bonafide void. 
If we imagine there is a spherical void with radius 11$h^{-1}$ Mpc lying only half in the survey, 
VoidFinder would be unable to fit a 10$h^{-1}$ Mpc sphere within the survey in this region, 
so any galaxies in the spherical void would not be identified as void galaxies. 
Galaxies living all along the boundaries of the SDSS mask could be 
affected by such misclassifications; therefore, we remove these galaxies 
from our analysis so as not to contaminate our wall sample. 
Even with the removal of these edge galaxies, we are still sampling a cosmologically significant volume.

We encourage the reader to keep in mind that when we refer to ``void'' and ``wall'' galaxies in this paper, 
the names are not synonymous with ``void'' and ``wall'' galaxies referred to in other papers that have 
utilized VoidFinder as a means of identifying large-scale voids. 
The classification of voids and walls is the same as seen elsewhere, but the samples of galaxies 
within those environments differs because of the differences in \HI-selected vs. optically-selected samples. 
In this paper, we are identifying the void and wall galaxies of a blind \HI\ survey. The galaxy sample 
used here and those used elsewhere are very different; for example, blind \HI\ detections residing in walls 
rarely populate the densest regions of the walls where we find a proliferation of galaxies in the optical. 
Optically-selected wall samples typically consist of galaxies from the red sequence. 
By using an \HI-selected sample, we lose a significant portion of the red sequence, and thus 
reduce the raw count of the wall population. This reduction of the wall population, results in an 
increased void fraction for \HI-selected samples. 
For comparison, within the same volume ($z \le 0.5$), $\sim26$ per cent of SDSS DR7 galaxies reside in voids.
Note that one must be careful in computing the void fraction of a sample. Calculating the numerator is 
straight forward and is defined as simply the number of objects residing in voids; while the 
denominator must include only the galaxies accessible to VoidFinder. Excluded from the denominator 
are galaxies beyond a specified redshift (here, $z \le 0.5$) and galaxies lying very near to the survey boundaries 
as discussed above. 

\begin{figure}
  \begin{center}
    \includegraphics[scale=0.2]{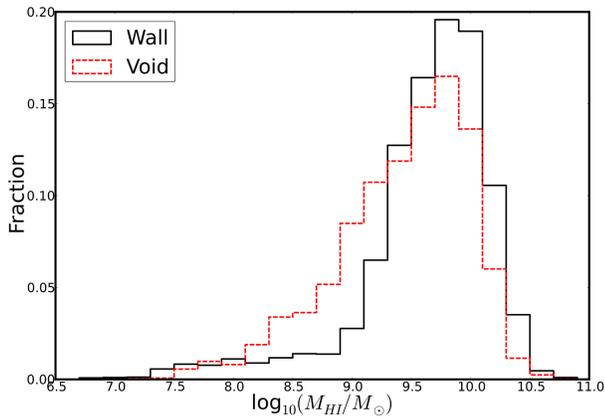} 
  \end{center}
  \caption{\small 
  \HI\ Mass distribution of void (red dashed line) and wall (black solid line) galaxies. 
  The relatively small number of nearby structures makes it imperative that we use an 
  inhomogeneity-independent method to estimate the \HI\ mass function of this dataset. 
  \HI\ detections living in voids typically have lower \HI\ masses than those in denser regions. 
    \normalsize}
  \label{fig:void_percentage}
\end{figure}

\begin{figure}
 \begin{center}
   \includegraphics[scale=0.19]{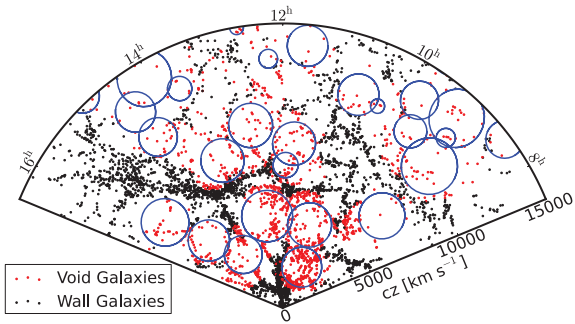} 
 \end{center}
 \caption{\small 10$h^{-1}$Mpc slab of ALFALFA detections in the Spring Sky.
 Void galaxies are displayed as red points, wall galaxies are displayed 
  as black points, and galaxies too close to the SDSS mask edge were not plotted. 
  The circles depict the intersection of the maximal sphere of a 
  void with the centre of the slice. The voids appear highly populated, but the galaxies 
  corresponding to the \HI\ detections are fainter than the $M_r<-20.09$ galaxies used in 
  our volume-limited catalog. Refer to Figure \ref{fig:vollimslice} for the volume-limited 
  sample used to produce the voids this slice. 
   \normalsize}
 \label{fig:slab}
\end{figure}

We present the distribution of \HI\ masses of void and wall galaxies of  
our ALFALFA sample in Figure \ref{fig:void_percentage}. Here, we notice void galaxies tend to have lower \HI\ 
masses than wall galaxies. 
Figure \ref{fig:slab} shows our ALFALFA sample in a 10$h^{-1}$ Mpc thick redshift slice 
centred at R.A.=12$^h$, Dec=10$^{\circ}$. This is the same slice depicted in 
Figure \ref{fig:vollimslice} where, again, wall galaxies are shown as black points, 
void galaxies are shown as red points, and circles depict the intersection of the 
maximal spheres of each void with the centre of the slice. The nearby voids appear highly populated 
by ALFALFA galaxies. The reader should keep in mind that 
Figure \ref{fig:slab} plots an \HI\ flux limited sample, whereas Figure \ref{fig:vollimslice} 
depicts an optical volume-limited sample; 
thus the majority of galaxies corresponding to these \HI\ detections found in voids 
are fainter than our volume-limited cut of $M_r<-20.09$. 

\begin{figure}
  \begin{center}
    \includegraphics[scale=0.2]{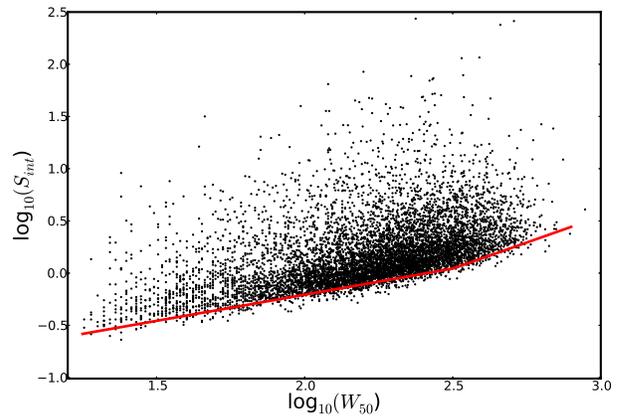} 
  \end{center}
  \caption{\small 
  Distribution of \HI\ detections used in this paper shown in integrated flux vs. velocity width space. The red line indicates our 
  adopted completeness limit which is the 50 per cent completeness threshold reported in Haynes et al. 2011. 
    \normalsize}
  \label{fig:Sint_w50_scatter}
\end{figure}

\begin{figure}
  \begin{center}
    \includegraphics[scale=0.2]{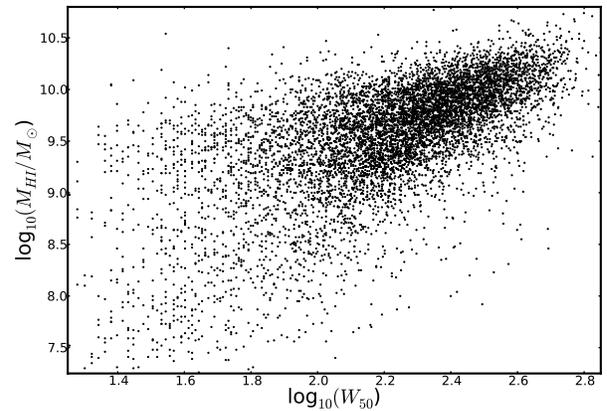} 
  \end{center}
  \caption{\small 
  Distribution of our final sample in \HI\ mass vs. velocity width space after all cuts have been made. 
  The completeness limit of the sample in this space is dependent on the distance to each object and 
  cannot be depicted using a single line.
    \normalsize}
  \label{fig:mass_w50_scatter}
\end{figure}

In addition to these physical boundary cuts, 
the method that we apply in this paper, the two dimensional stepwise 
maximum likelihood (2DSWML) method, requires that our sample be  
complete; therefore, we eliminate all galaxies that fall below 
the 50 per cent completeness threshold in the flux-width plane reported 
in Section 6 of \cite{Haynes2011}. 
Figure \ref{fig:Sint_w50_scatter} depicts the distribution of \HI\ detections described in Section \ref{subsec:ALFALFAdata} with 
this completeness threshold in the flux-width plane.
This cut eliminates 152 void galaxies and 240 wall galaxies.
We provide a brief explanation of the completeness cuts here. 
\cite{Haynes2011} begin by separating the ALFALFA detections into \HI\ line-width bins, 
then within each width bin the galaxies are binned again by integrated \HI\ flux. 
A flux-limited sample from a uniform distribution of galaxies 
will produce a number count that approximately follows a power law with slope -3/2. The authors 
determine the onset of incompleteness when the binned integrated flux data deviate from this form. 
In Figure \ref{fig:mass_w50_scatter}, we present the distribution of the final sample used in our analysis 
in the mass-width plane.

We make cuts to the ALFALFA-SDSS cross-referenced catalog identical to the ones mentioned in Section \ref{subsec:ALFALFAdata}.
The resulting subsample is 305 \HI\ detections fewer than the full sample mentioned above. 
We lose 54 \HI\ detections from the void sample, 132 detections from the wall sample, and 
119 from the edges where the survey region is inaccessible to VoidFinder.
We further limit this sample to objects lying only in the ALFALFA Spring Sky. 
These reductions leave us with 7404 \HI\ detections. 2571 of these live in voids, while 4485 reside in denser regions. 
The remainder of the galaxies lie along the edge of the SDSS DR7 survey mask in regions inaccessible to 
VoidFinder. 

\subsection{The 2DSWML Method}
\label{subsec:Method}
Because the ALFALFA survey's detection limit is dependent on both 
\HI\ mass and velocity width \citep{ALFALFAI}, 
we estimate the \HI\ mass function (HIMF) and the velocity width function (WF) 
of our ALFALFA void and wall samples using 
an extension of the stepwise maximum likelihood method of \cite{EEP1988} 
called the bivariate or two-dimensional stepwise maximum likelihood (2DSWML) method, 
introduced by \cite{Loveday2000}. We provide a brief overview of the 2DSWML method 
here. For more details of the direct application of this method 
to the $\alpha$.40 sample, see Martin et al. (2010, Appendix B).

We obtain our estimates of the HIMF and WF by first splitting the 
bivariate distribution function $\phi(M_{\rm HI},W_{50})$ into bins and 
define $\phi_{jk}$ as the maximum likelihood parameter of the 
distribution in logarithmic mass bin $j$ and
logarithmic velocity width bin $k$. 
We attain the maximum likelihood solution via  
iteration of the following equation obtained by
\cite{EEP1988}: 
\begin{equation}
\phi_{jk}= \frac{n_{jk}}{\sum_i\frac{H_{ijk}}{\sum_m\sum_nH_{imn}\phi_{mn}}}.
\label{eq:iterationEquation}
\end{equation}
Here, $n_{jk}$ is the number of galaxies within each bin in the
mass-width plane and $H_{ijk}$ is a function that ensures the
summation only goes over the area of bins accessible to galaxy $i$, where
\begin{equation}
H_{ijk}= \frac{1}{\delta M_{\rm HI} \delta W_{50}}\int_{W_k^-}^{W_k^+}\int_{M_j^-}^{M_j^+} C_i(M_{\rm HI},W_{50})dM_{\rm HI}dW_{50}.
\label{eq:Hfunction}
\end{equation}
Here $C_i(M_{\rm HI},W_{50})$ is an approximation of the 
completeness function described in \citealt{Haynes2011} in the
mass-width plane for galaxies at distance $D_i$. 
That is, if an object's integrated flux falls below the completeness limit, $C$=0, otherwise $C$=1.  
$M_j^-$ and $M_j^+$ are the lower and upper limits on logarithmic mass bin $j$, and similarly,
$W_k^-$ and $W_k^+$ are the lower and upper limits on logarithmic width bin $k$.

After obtaining the maximum likelihood bivariate distribution parameters, $\phi_{jk}$, 
we marginalize over velocity width to measure the HIMF. Because the normalization 
is lost, we match the normalization of the void and wall galaxy HIMFs to the number density 
of void and wall galaxies within their respective volumes.
We compare our measurements of the void and wall galaxy HIMFs 
over the mass range $7.2<\log\left(M_{\rm HI}/M_{\odot}\right)<10.8$ 
with a Schechter function \citep{Schechter1976} of 
the form
\begin{eqnarray}
\Phi(M_{\rm HI}) = \frac{dN}{d\log M_{\rm HI}} =\ln 10 \Phi^* \left(\frac{M_{\rm HI}}{M^*}\right)^{(\alpha+1)}
\nonumber\\
\times\exp\left(-\frac{M_{\rm HI}}{M^*}\right).
\label{eq:Schechter}
\end{eqnarray} 
We estimate the normalization factor $\Phi^*$, the characteristic gas mass $M^{*}$, 
and the low-mass-end slope $\alpha$ using a least squares estimator.

Here we present the global HIMF of the full $\alpha.40$ data set. It is well fit by a Schechter function with estimated 
parameters ($\Phi^*=(6.3\pm0.3) \times 10^{-3}$Mpc$^{-3}$, $\log$($M^*/M_{\odot}$)+ 2log$h_{70}=9.96\pm0.02$, $\alpha=-1.33\pm0.02$) 
similar to those of \cite{Martin2010}:
$\Phi^*=(4.8\pm0.3) \times 10^{-3}$Mpc$^{-3}$, $\log$($M^*/M_{\odot}$)+ 2log$h_{70}=9.96\pm0.02$, $\alpha=-1.33\pm0.02$. 
Limiting the data set to \HI\ detections located in the Spring Sky, 
we do not see a significant difference in the best-fitting parameters of the HIMF 
($\Phi^*=(5.34\pm0.4) \times 10^{-3}$Mpc$^{-3}$, $\log$($M^*/M_{\odot}$)+ 2log$h_{70}=9.97\pm0.02$, $\alpha=-1.35\pm0.04$).

\begin{figure}
  \begin{center}
    \includegraphics[scale=0.165,trim=3.25cm 1cm 5cm 2cm, clip=true]{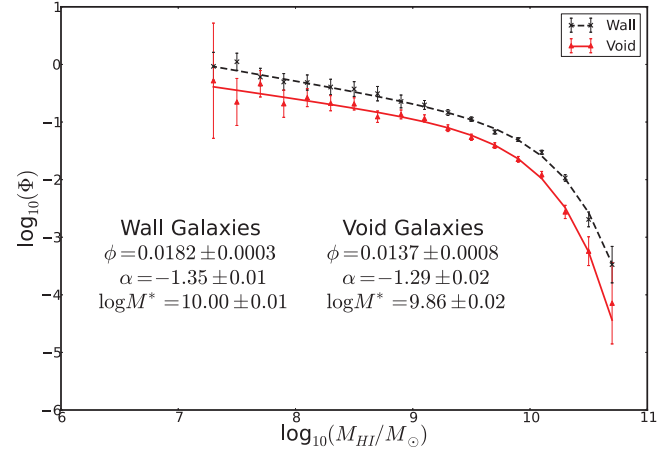}
  \end{center}
  \caption{\small 
  The HIMF of void (red) and wall (black) galaxies. 
  The best-fitting Schechter functions for each sample are shown as solid (void) and dashed (wall) lines. 
  We notice a shift towards lower masses as well as a small decrease in the 
  low-mass slope for void galaxies. 
    \normalsize}
  \label{fig:void_himf}
\end{figure}

Dividing the Spring Sky into void and wall galaxies produce the following results:
In Figure \ref{fig:void_himf} we present the HIMF of both void and wall galaxy samples. 
For the void sample, we estimate the best-fitting Schechter parameters 
to be 
$\Phi^*=(1.37\pm0.1) \times 10^{-2}$Mpc$^{-3}$, $\log$($M^*/M_{\odot}$)+ 2log$h_{70}=9.86\pm0.02$, $\alpha=-1.29\pm0.02$.
For the wall galaxy sample, we estimate 
$\Phi^*=(1.82\pm0.03) \times 10^{-2}$Mpc$^{-3}$, $\log$($M^*/M_{\odot}$)+ 2log$h_{70}=10.00\pm0.01$, $\alpha=-1.35\pm0.01$.
See Section \ref{subsec:Errors} for an explanation of uncertainties. 
The curves in Figure \ref{fig:void_himf} show the Schechter functions associated with 
these best-fitting parameters. 

We see that  $\log(M^*)$ shifts towards lower \HI\ masses 
in voids by $\log(M^*)=0.14$. This characteristic mass is shifted by a factor of 1.4 and the direction of this shift is consistent with 
the shift in the DMH mass function of extended Press-Schechter theory \citep{Goldberg2005}. 
This direction of the shift is also in agreement with the shift in the optical LF. We cannot quantitatively compare 
our HIMF shift with the shift found in the optically-selected LF from voids to walls because of a difference in sample 
selection, but for completeness, \cite{Hoyle2005} find 
a shift in LF of the $r$-band magnitude towards lower magnitudes of $M_r^*\sim 1$ for void galaxies (a factor of 2.5). 
We also see only a slight dependence of the low-mass slope, $\alpha$, on environment, with the slope 
steepening with increasing density. 
We compare our results to previous work more completely later in Section \ref{sec:discuss}, 
but for now we will briefly compare our HIMF resutls to 
the HIMF of the Leo Group \citep{Stierwalt2009},
who estimate a low-mass slope of $\alpha=-1.41\pm0.04$ using ALFALFA.
Given our findings,
we conclude that the low-mass slope of the HIMF for ALFALFA galaxies in the Spring Sky is shallow 
for voids. The slope may increase with density to arrive at the steep Leo I Group slope, but 
the slope does not necessarily increase monotonically with density.   
Depending on how densely packed a particular galactic group or cluster is, 
the low-mass slope of the HIMF may vary drastically, ranging anywhere from the 
steep slope of the Leo Group \citep{Stierwalt2009} to the very flat slopes found in 
loose groups \citep{Pisano2011} and clusters
\citep{Verheijen2001,RosenbergSchneider2002,Martin2011} which, when combined, 
may give an intermediate low-mass slope similar to that of our wall galaxy sample.
 
\subsubsection{WF}
\label{subsubsec:WF}
To determine the velocity width function, we obtain our results 
from the bivariate distribution function of the 2DSWML method and
marginalize over \HI\ mass. 
As with the HIMFs, we must match the normalization of the 
void and wall WFs to their respective densities.
To remain consistent with previously reported ALFALFA WF results 
(\citealt{Papastergis2011} obtain observational results and translate theoretical results 
to the observed quantity rather than estimate corrections for the observations), 
we have corrected the velocity widths for Doppler broadening and make 
no other corrections, e.g. galaxy inclination. 
We compare the void and wall galaxy width functions over the 
width range $1.3\log($km s$^{-1})<\log\left(W_{50}\right)<2.9\log($kms$^{-1})$, 
where $W_{50}$ is the velocity width measured at 50 per cent of the peaks of the \HI-line profile. 
The high velocity width end falls off too quickly to be well fit 
by a Schechter function, so we fit the WFs to a modified Schechter function 
of the form 
\begin{equation}
\Phi(W_{50}) = \frac{dN}{d\log W_{50}} = \ln 10 \Phi^* \left(\frac{W_{50}}{W^*}\right)^{\alpha}\exp\left(-\frac{W_{50}}{W^*}\right)^{\beta}
\label{eq:modifiedSchechter}
\end{equation}
and estimate the 
normalization factor $\Phi^*$, the characteristic velocity width $\log W^{*}$, 
and the low and high velocity width slopes $\alpha$ and $\beta$ 
using a least squares estimator.

\begin{figure}
 \begin{center}
    \includegraphics[scale=0.165,trim=2.85cm 1cm 5cm 2cm, clip=true]{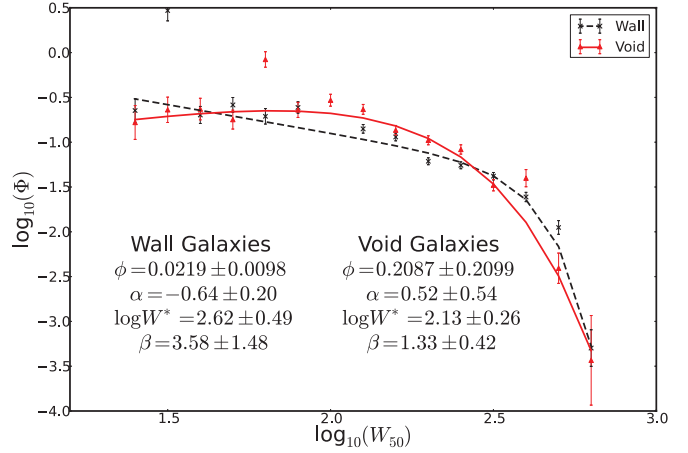} 
  \end{center}
  \caption{\small 
  The WF of void (red) and wall (black) galaxies. 
  The best-fitting modified Schechter functions for each sample are shown as solid (void) and dashed (wall) lines. 
  Neither the void nor wall WF is well fit by a modified Schechter function.
    \normalsize}
  \label{fig:void_wf}
\end{figure}

Here we present the global WF of the full $\alpha.40$ sample. It is well fit by a modified Schechter function where 
our 2DSWML-estimated parameters 
($\Phi^*=(2.1\pm0.2) \times 10^{-2}$Mpc$^{-3}$, $\log(W^*)=2.56\pm0.03$, $\alpha=-0.73\pm0.02$, $\beta=2.6\pm0.2$) 
match closely those of \cite{Papastergis2011}: 
$\Phi^*=(1.1\pm0.2) \times 10^{-2}$Mpc$^{-3}$, $\log(W^*)=2.58\pm0.03$, $\alpha=-0.68\pm0.11$, $\beta=2.7\pm0.3$. 
Limiting the data set to \HI\ detections located in the Spring Sky, 
the WF remains well fit by a modified Schechter function although the parameters change somewhat: 
$\Phi^*=(1.3\pm0.3) \times 10^{-2}$Mpc$^{-3}$, $\log(W^*)=2.61\pm0.03$, $\alpha=-0.91\pm0.02$, $\beta=3.44\pm0.2$.
When we divide the Spring Sky sample into void and wall galaxies, 
our results of the 2DSWML method change drastically.
In Figure \ref{fig:void_wf} we present the WF of both void and wall galaxy samples. 
We see these functions are not as well fit by the modified Schechter function as the 
global $\alpha.40$ WF (see Figure 4 in \citealt{Papastergis2011}). 
For the void sample, we estimate the best-fitting modified Schechter parameters 
to be 
$\Phi^*=0.21\pm0.1 $Mpc$^{-3}$, $\log(W^*)=2.13\pm0.3$, $\alpha=0.52\pm0.5$, $\beta=1.3\pm0.4$.
For the wall galaxy sample, we estimate 
$\Phi^*=0.022\pm0.009 $Mpc$^{-3}$, $\log(W^*)=2.62\pm0.5$, $\alpha=-0.64\pm0.2$, $\beta=3.58\pm1.5$.
We note here that two bins in both the void and wall WFs appear 
to be extreme outliers (void: $\log(W_{50})=1.8, 2.6$; wall: $\log(W_{50})=1.5, 2.0$), however 
these bin heights are not caused by low-number statistics.
The curves in Figure \ref{fig:void_wf} show the modified Schechter functions associated with 
these best-fitting parameters.

\begin{figure}
  \begin{center}
    \includegraphics[scale=0.45]{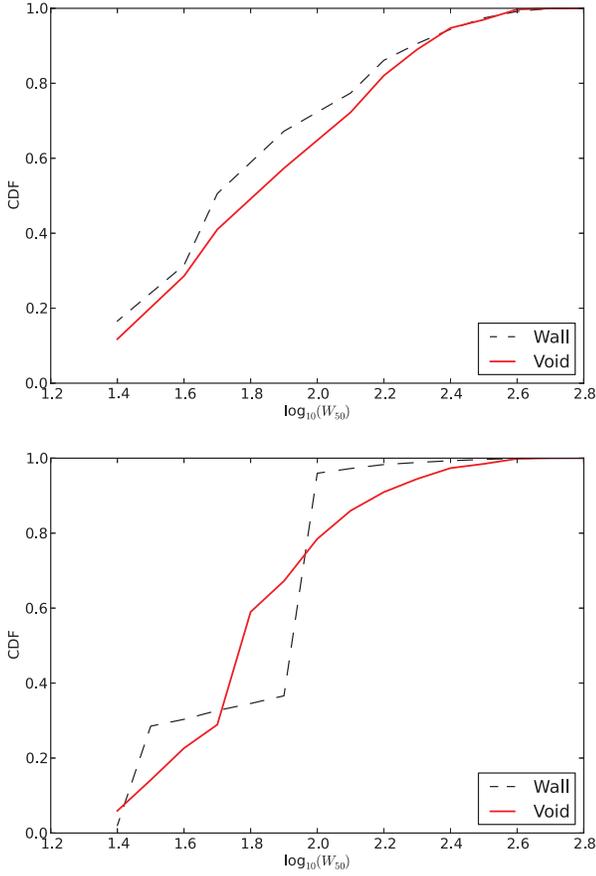}
  \end{center}
  \caption{\small 
  Cumulative distribution function of wall (black dashed) and void (red solid) WF.
  Top: Outlying points from the distributions have been removed. 
  Bottom: Outlying distribution points are left in the CDF calculation.
    \normalsize}
  \label{fig:ks_test_WF} 
\end{figure}

We see that $W^*$ shifts towards lower velocity widths 
in voids compared to walls; however, due to large uncertainties, we are unable to make any
conclusive statements about the other parameters. 
It is our goal to ascertain whether there are any differences in the WF of void galaxies and wall galaxies.
Looking at the poorly-fit Schechter functions gives us no real insight, so
we compute a Kolmogorov--Smirnov test to compare the wall and void WF distributions. 
Including all points in the distributions (note the outlying points), we obtain a p-value of 0.05
from the K-S test. That is, 95 per cent of the time we would correctly reject the hypothesis that 
the void and wall WFs are drawn from the same distribution. 
We have no reason to exclude the extreme outliers, but excluding the extreme outliers, yields a p-value of 0.07. 
See Figure \ref{fig:ks_test_WF} for a comparison of the CDFs of the two distributions with 
and without the outliers.
\begin{figure}
  \begin{center}
    \includegraphics[scale=0.165,trim=3.25cm 1cm 5cm 2cm, clip=true]{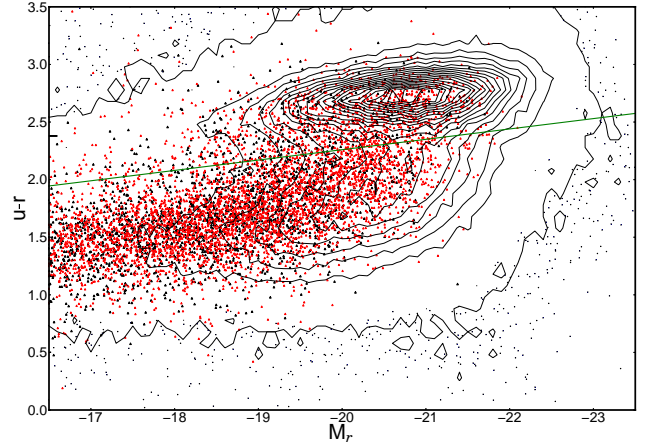} 
  \end{center}
  \caption{\small 
  Colour-magnitude diagram of optically-selected and \HI-selected galaxies. 
  Contours and black outlying points represent density of SDSS DR7 galaxies. Black and red triangles represent 
  ALFALFA wall and void galaxies respectively. Green line shows the colour cut mentioned in the text that splits the samples 
  into a ``red sequence''  and a ``blue cloud''. 
  The ALFALFA sample is lacking in the red sequence where we find our optical sample to be most dense. 
    \normalsize}
  \label{fig:ur_mag_contour_with_color_line} 
\end{figure}
\begin{figure}
  \begin{center}
    \includegraphics[scale=0.165,trim=3.25cm 1cm 5cm 2cm, clip=true]{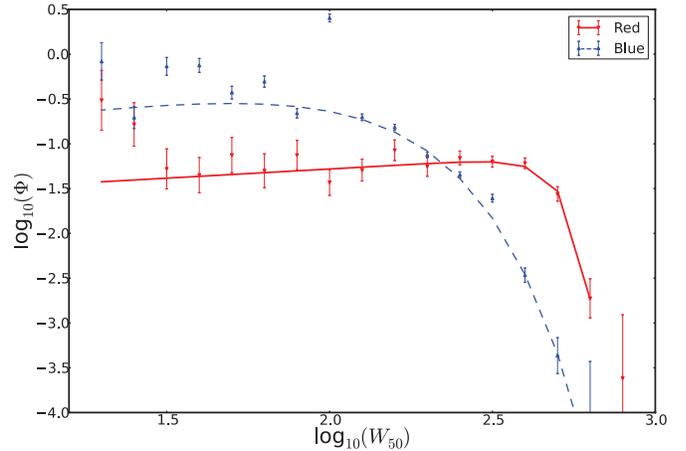}
  \end{center}
  \caption{\small 
    The WF of red (red triangles) and blue (blue triangles) galaxies. 
  The best-fitting modified Schechter functions for each sample are shown as solid (red) and dashed (blue) lines. 
  We see that the low-width slope of the red galaxy WF is substantially flatter than that of the blue sample. 
    \normalsize}
  \label{fig:wf_blue_red}
\end{figure}

As mentioned above, galaxies in voids tend to be blue, late-type galaxies whereas red, dead, elliptical 
galaxies are primarily found in denser regions of the Universe. 
Therefore, we investigate the effects of colour to shed some light on 
possible explanations as to why/if the WF 
differs between void and wall galaxies.
For this sample, we use the ALFALFA-SDSS cross-reference catalog provided by \cite{Haynes2011}. 
This results in a few less galaxies than the sample we have been working with because we are dependent on the 
SDSS photometric pipeline for matches, and in some cases, the galaxy images may be contaminated 
by a foreground star or other artefact. 

We split our \HI\ cloud sample into groups based on colour where we define 
galaxies with a colour $u-r<-0.09 M_r+0.46$ as blue and galaxies with colour $u-r\ge-0.09M_r+0.46$ as red. 
Here, $M_r$ is the $r$-band magnitude of the object's optical counterpart given by the SDSS DR7. 
See Figure \ref{fig:ur_mag_contour_with_color_line} for the location of this definition 
of colour as shown on
a colour-magnitude diagram; this cut divides the sample into a ``red sequence'' and a ``blue cloud''.
Our resulting subsamples contain $\sim 6,000$ blue galaxies and $\sim 1,400$ red galaxies.
Figure \ref{fig:wf_blue_red} depicts the resulting WFs of the blue and red subsamples along with the best-fitting 
modified Schechter functions. 
We see that the low-width slope of the red galaxy WF is substantially 
flatter than that of the blue galaxies.
Because a blind-\HI\ survey is more likely to identify a blue, late-type galaxy (e.g. \citealt{Huang2012}), 
it is understandable that the low-width slope of the red galaxy WF would be 
flat or even turn over,
because only the most massive red elliptical galaxies 
will contain enough \HI\ to be detected, therefore boosting the high-width end.

\subsection{Error Analysis}
\label{subsec:Errors}
We account for a number of sources of error which we add in quadrature; the first 
is from Poisson counting errors. 
We also take into account the error in each bin of our bivariate distribution 
introduced from the 2DSWML 
method which we estimate via the inverse of the information matrix
as described in \cite{EEP1988}. This method uses the fact 
that the maximum likelihood estimates of $\phi_{jk}$ are asymptotically 
normally distributed.

We also account for cosmic variance, or the effects of the 
inhomogeneity of large scale structure, 
by using the jackknife resampling method \citep{Efron1982} 
to estimate the uncertainties in our HIMF/WF measurements. We do so by dividing 
the Spring Sky into $N$ subregions, and calculating the HIMF/WF of a subsample 
of galaxies which omits a different 1/$N$th of the surveyed area each time. 
After measuring the HIMF/WF of each of the $N$ regions, we estimate the 
variance of the value of the distribution function in each \HI\ mass/width bin as 
\begin{equation}
\mbox{Var} (x) = \frac{N-1}{N} \sum_{i=1}^{N}\left(x_{i}-\bar{x}\right)^{2}.
\label{eq:variance}
\end{equation}
For our analysis, we divide our sample volume into $N=18$  subregions, equally spaced in R.A.

For the HIMF, we need to account for errors in distance measurements. 
The calculation of the \HI\ mass of a galaxy is 
proportional to the square of the distance to the galaxy; thus, errors 
in distance estimates could lead to large errors in mass and cause 
a galaxy to be moved from one mass bin to another. 
We account for this by creating 300 realizations 
of the void and wall HIMFs, where we estimate errors 
on $M_{\rm HI}$ using reported $S_{int}$ errors and estimated distance errors. 
We give each galaxy $i$ in our sample, at a distance $D_i$, a random \HI\ mass assuming 
a Gaussian random error, and calculate the HIMF of each of these realizations.

We use a similar approach to account for the error in velocity widths which can 
shift a galaxy from one width bin to another, by creating 150 realizations of the 
void and wall WFs using the reported $W_{50}$  and $\Delta W_{50}$ values. 
The published $W_{50}$ errors, $\Delta W_{50}$, from the $\alpha.40$ catalog 
take into account  distortion of the spectral profile due to noise and 
the systematic effect of a guess at the locations of the spectral boundaries.

\section{Comparison with Previous Observations}
\label{sec:discuss}
\subsection{Comparing the HIMF}
\label{subsec:himf_comp}
Previous studies on the environmental impact of the HIMF come to varying conclusions 
depending on the \HI\ sample used or the method of determining galaxy density.
\cite{Springob2005} compare the HIMF of a complete optically-selected sample 
across environments. The sample consists of \HI-selected galaxies with integrated \HI\ flux $S_{int}>0.6$
within $-2^{\circ}<$Dec$<38^{\circ}$ with an optical apparent diameter 
$a>1.'0$, Galactic latitude $|b|>15^{\circ}$, and morphological type Sa-Irr. 
The authors define environment based on ranges of galaxy
densities (local density: $n<1.5, 1.5<n<3.0, n>3.0$) 
defined via a Point Source Catalog Redshift Survey (PSCz) density reconstruction map, 
and find, with low-significance, that $\alpha$ flattens with increasing 
density while $M^*$ shifts to lower masses (within a range of $\log(M^*)<0.16$). 
This trend is consistent with the 
theory that galaxies in clusters are \HI\ deficient (Giovanelli \& Haynes 1985b), 
however the decrease in $M^*$ with increasing density contradicts our findings.
While these findings are of low statistical significance, 
and the authors report that better statistics are needed
to conclusively determine whether the HIMF varies across environments,
differences between our results and theirs could be due to the difference in 
samples -- optically-selected vs. \HI-selected samples. The authors also do not 
isolate ``void" galaxies as we have defined them. 

\cite{Zwaan2005} compute the HIMF in several density regions  using 
$n$-th nearest neighbor distances to define density for $n=1,3,5,10$.
They find that the best-fitting Schechter function for low-density environments has a 
flat low-mass end slope $\alpha$ that steepens with increasing density when 
defined on small and large scales. They find no environmental dependence 
of the characteristic turnover mass $M^*$. 
This is counterintuitive given that we would expect galaxies in extremely overdense 
areas to have their gas stripped away during mergers or other galactic interactions.
One possible reason for \cite{Zwaan2005} finding 
such a contrasting result is that the authors define local density using the 
\HI\ galaxies themselves via nearest neighbor distances. 
Galaxies living in extremely high-density environments defined by an optical 
sample may undergo mergers or other astrophysical effects causing most, if not all, 
\HI\ to be stripped from low mass galaxies, resulting in a very different density map 
from those found using optical surveys.

\cite{Martin2011} uses both the full $\alpha$.40 dataset as well as the Spring Sky 
portion of the data to estimate the HIMF of galaxies living in several different 
density regions. She uses two methods to determine local density: first by using 
two PSCz density reconstruction maps (a coarse and fine grid), and second by applying an 
$n$-th nearest neighbors method to a volume-limited sample of 
galaxies from SDSS DR7 over a range of $n$ values, and thus, over a range of density scales.
The latter of these methods 
was only applied to the Spring Sky portion of the dataset where the $\alpha$.40 
and SDSS DR7 footprints overlap.
Using these methods, Martin (2011) finds that $M^*$ increases with 
increasing density on all scales, whereas the trend of the low-mass-end slope 
varies depending on the scale on which environment is defined. 
When density is defined on larger scales, $\alpha$ steepens with density 
until some threshold density, after which it flattens out.
The author suggests that 
when density is defined
on the largest scales, the HIMF will have a flattened 
low-mass slope, in the extreme high and low density regions, 
whereas the not-so-extreme environments will have a relatively 
steeper slope that increases with density.
\cite{Martin2011} proposes the lowest density regions maintain an abundance of intermediate-mass 
galaxies -- possibly due to an increased sensitivity to reionization in voids \citep{Hoeft2006} -- which 
boosts the intermediate regions of the HIMF thus flattening the slope. In the highest density regions, 
the lowest mass galaxies have their cool gas stripped away, thus flattening the low-mass slope. 

\cite{Pisano2011} use the HICAT \citep{Meyer2004} catalog and the Lyons Group Galaxies to 
determine the HIMF of six loose groups of galaxies 
with properties similar to those of the Local Group. 
These authors find a very flat low mass slope for loose galactic groups. 
\cite{Stierwalt2009} use the ALFALFA survey calculate the HIMF of the 
Leo I Group which has a higher 
density than the groups used in \citealt{Pisano2011}. 
They find a steep low-mass slope of $\alpha=-1.41$.
\citep{Kovac2007PhD} on the other hand, find a very flat low-mass slope of the high-density 
CV group. There does not seem to be a clear trend with environment at least on the group level.

In the Spring Sky, we investigate the HIMF parameters of true 
voids -- extreme underdensities on the largest scales -- and 
find the characteristic turnover mass $M^*$ is shifted towards lower masses
compared to $M^*$ in denser environments. 
In conjunction with the findings of \cite{Stierwalt2009}, who 
find a relatively steep ($\alpha=-1.41$) low-mass-end of the 
HIMF for objects in the Leo I group, our findings of the low-mass slope of void galaxies 
($\alpha=-1.29$) indicate that the low-mass slope of the HIMF of ALFALFA detections steepens with density. 
This result is in agreement with the findings of Martin (2011). 
A number of things could be responsible for the somewhat flatter low-mass slope of void galaxies than wall galaxies. 
The higher specific star formation rates of low mass galaxies in voids may cause them to 
burn their fuel faster than their counterparts in walls, thereby flattening the low-mass slope. 
Given the shift in characteristic \HI\ mass to lower masses in voids, and 
assuming (again, naively) that the baryonic mass is directly proportional to the DMH mass,
void galaxies will have shallower 
potential wells and could much more easily lose their cool gas, causing their low-mass slope to flatten.  
Another, more likely, candidate for the cause of the flattened low-mass slope is supernova feedback. 
While a comparison of optical and infrared redshift 
surveys \citep{Fisher1995,saunders2000,Strauss2002,Jones2004} might reveal 
similar large-scale structure \citep{HoyleVogeley2004}, 
local density maps revealing large-scale structure in \HI\ would differ 
drastically (e.g. \cite{Papastergis2013clustering,Waugh2002}).
Therefore, we cannot quantitatively compare our results to \cite{Zwaan2005}, because they 
measure local density using an \HI\ sample.
We also cannot quantitatively compare our results to those of \cite{Springob2005}, because they 
determine the HIMF using an optically-selected sample.

\begin{table*}
\begin{centering}
\begin{tabular}{lllcccc}
Author &Sample &Density& $\Phi^* $  & $M_{\rm r}^*$ + 2log$h$ & $\alpha$ &  \\ 
 & &Measured & $\times 10^{-2}h^3$Mpc$^{-3}$ & & & \\ \hline

Rosenberg  &ADBS& Full Sample&  0.58 & 9.88 & -1.53  &  \\ 
 && Virgo Cluster &   &  & -1.2  &  \\ \hline 

Springob  &optically-selected& PSCz DM: -lowest $\rho$ & 0.32 & 10.07 & $\sim$-1.37  &  \\
  & galaxies & -intermediate $\rho$&  & 9.91  & $\sim$-1.13  &  \\ 
  &  & -highest $\rho$&  & 9.96  & $\sim$-1.25  &  \\ \hline 

Zwaan  &HIPASS& \HI\ $10^{th}$NN: -lowest $\rho$&  0.60 & 9.80 & -1.0  &  \\
  && -intermediate $\rho$  & & & -1.4  &  \\   
  & & -highest $\rho$&   & & -1.55  &  \\ \hline 

Martin  &ALFALFA& SDSS $10^{th}$NN -lowest $\rho$&   0.31& 9.80 & -1.15 &  \\ 
  && -intermediate $\rho$  &    & 10.05 & -1.37  &  \\   
  && -highest $\rho$  &    & 10.05 & -1.2  &  \\  \hline 
Martin  &ALFALFA& Coarse PSCz DM: -lowest $\rho$&   & 9.93 & -1.0 & \\  
  && -higher $\rho$  &    & 10.05 & -1.45  &  \\   \hline
Stierwalt &ALFALFA & Leo I Group&  $\sim$3.0 & $\sim$10.7 & -1.41& \\ 
 & & Full Sample&  & & & \\ \hline 
Pisano &HICAT & Optically-Selected&  & $\sim$9.8 & -1.0& \\ 
 & & Loose Groups&  & & & \\ \hline 

this paper &ALFALFA & VoidFinder - Void&  1.37$\pm0.08$ & 9.86$\pm0.02$ & -1.29$\pm 0.02$ & \\ 

 & & VoidFinder - Wall&  0.28$\pm0.03$ & 10.00$\pm0.03$ & -1.35$\pm 0.03$ & \\ 

\end{tabular}
\caption{Schechter function fits to the HIMFs of different galaxy samples across environments.
Each sample contains either \HI-selected or optically-selected galaxies, and 
galaxy environment was determined using either Nearest Neighbor algorithms (using optical or \HI\ samples), 
PSCz density maps, well-known groups/clusters, or VoidFinder.}
\label{tab:values}
\end{centering}
\end{table*}

\subsection{Comparing the WF}
\label{subsec:wf_comp}

Little work has been done to 
determine the environmental effects on purely 
observed velocity width functions, specifically in the most underdense regimes.
\cite{Desai2004} compare the Galaxy Circular Velocity Function (GCVF) of clusters 
grown in a $\Lambda$CDM simulation with the 
GCVF of clusters from the SDSS obtained using galaxy photometry and 
the Tully-Fisher relation (Tully \& Fisher 1977). They find that both the 
observed and simulated GCVFs are well fit by a power law, with the observed 
cluster GCVF having only a slightly shallower slope, 
although they mention the difference in slope is not significant. 
They also determine the GCVF of ``field'' galaxies and find that, compared 
to the predicted power law, these galaxies display a much shallower 
slope at low velocities and a much steeper slope at higher velocities with a 
turnover velocity of $\sim 200$km s$^{-1}$ -- a shape well described by a Schechter function. 

\cite{Abramson2013} report their findings on the 
circular velocity function (CVF) of galaxies in groups 
using three different samples from the 
NYU Value Added Galaxy Catalog \citep{Blanton2005NYUVAGC}.
In contrast to $\Lambda$CDM predictions, 
these authors find that the group galaxy CVF is consistent with, 
if not shallower than, the 
observed field galaxy CVF at the low-velocity end. 
They find that the shape of the CVF depends primarily on 
the morphological types of galaxies included in the samples, with an increase in 
the fraction of late-type galaxies steepening the low-velocity slope; they suggest the 
flattened slope of the group CVF is due primarily to the depression of late-type galaxies in groups.
\cite{Sigad2000} find similar results using a 
$\Lambda$CDM simulation. Assuming a linear relation between halo mass 
and luminosity, the authors populate the haloes with ``galaxies", identify group and 
isolated ``galaxies", and report 
that the isolated ``galaxies" have a steeper low-velocity slope than grouped galaxies. 

Although we cannot directly compare an observed WF with a CVF, 
at first glance, 
our findings on the environmental dependence of the WF appear to be 
in contrast with \cite{Abramson2013} and \cite{Sigad2000}, 
and indicate that galaxies in 
voids exhibit a shallower low-velocity slope than their counterparts 
in denser regions; however, large uncertainties weaken our conclusions. 
On the other hand, assuming colour and morphology go hand-in-hand, our results are in agreement with 
those of \cite{Abramson2013} in that blue galaxies seem to have a steep 
 low-velocity width slope and red galaxies tend to have a much shallower low-velocity width slope. 
See Figure \ref{fig:wf_blue_red} for a comparison of the blue and red WFs. 

\section{Conclusions}
\label{sec:conc}

Using the void catalog obtained by \cite{Pan2012} and the
$\alpha.40$ catalog from \cite{Haynes2011}, we measure the HIMF of
2,300 void galaxies with \HI\ masses ranging from $7.2<\log\left(M_{\rm HI}/M_{\odot}\right)<10.8$.
We find that the HIMF of void galaxies is well fit by a Schechter
function with parameters 
$\Phi^*=(1.37\pm0.1) \times 10^{-2} h_{70}^3$Mpc$^{-3}$, 
$M_{\rm HI}^*$+ 2$\log h_{70}=9.86\pm0.02$, and $\alpha=-1.29\pm0.02$.
For galaxies residing in higher density regions, we find the best-fitting Schechter parameters
to be $\Phi^*=(1.82\pm0.03) \times 10^{-2} h_{70}^3$Mpc$^{-3}$, 
$M_{\rm HI}^*$+ 2$\log h_{70}=10.00\pm0.01$, and $\alpha=-1.35\pm0.01$.
We also measure the WFs across environments and, 
while the void and wall WFs are not well fit by a modified Schechter function, we estimate the 
best-fitting parameters of a modified Schechter function to be 
$\Phi^*=0.21\pm0.1 h^3$Mpc$^{-3}$, 
$\log(W^*)=2.13\pm0.3$, $\alpha=0.52\pm0.5$, and $\beta=1.3\pm0.4$ for the void sample, 
and 
$\Phi^*=0.022\pm0.009 h^3$Mpc$^{-3}$, 
$\log(W^*)=2.62\pm0.5$, $\alpha=-0.64\pm0.2$, and $\beta=3.58\pm1.5$ for the wall sample.

We conclude the following:

1. The ALFALFA $\alpha.40$ catalog yields a higher fraction of 
void galaxies (33\%) than the optically-selected (26\%) SDSS DR7 magnitude-limited sample
over the same volume. 
Because these surveys are magnitude limited, the fraction of void galaxies 
varies with redshift. We know blind-\HI\ surveys preferentially detect blue, late-type galaxies and 
void regions are also predominantly filled with these blue, spiral galaxies. The red luminous galaxies, 
typical in the wall sample of an optical survey, are lost when we use an \HI-selected sample. 
This reduction of red galaxies caused by moving from an optically-selected sample to an \HI-selected sample, 
therefore, decreases the raw count of the wall sample, and increases the void fraction for \HI-selected samples. 

2. Our findings suggest that the characteristic turnover mass of the HIMF is marginally dependent on environment. 
The characteristic \HI\ mass, $M^*$, shifts to lower masses in voids by a factor of 1.4; while this shift is small, it is significant. 
The shift is consistent with extended Press-Schechter theory which states that the mass function 
should shift to lower masses in underdense regions. 

3. We see only a slight difference in the low-mass slopes of void and wall 
galaxy HIMFs. 
We believe something may be gleaned from combining our void results with other's results  
investigating large-scale, non-void environments.
When we couple our void galaxy HIMF with other \HI-selected samples from optically-selected environments,
such as Stierwalt et al.'s (2009) HIMF of the Leo I group, 
and Rosenberg \& Schneider's HIMF of galaxies in the Virgo Cluster,
we find that the low-mass end slope $\alpha$ varies with environment on the largest scales.
This indicates a possible trend with environment (as mentioned in \citealt{Martin2011}) where the low-mass slope 
is flat in voids, and increases with density up to some turnover density, where the galaxies within 
clusters become \HI\ deficient through e.g. galaxy-galaxy interactions.

Our wall HIMF is neither as steep as that found 
in \cite{Stierwalt2009} nor as flat as 
the extremely overdense (cluster) HIMFs suggested by 
\cite{RosenbergSchneider2002} and \cite{Martin2011},
because we are effectively averaging over all non-void densities. 
Our wall regions cover a combination of high-density groups made up of early-type galaxies, 
low-density groups made up of late-type galaxies, 
and clusters which tend to be \HI\ deficient, just to name a few. 
This conglomeration of vastly different environments yields a wall HIMF with a low-mass slope 
somewhere inbetween 
the flattest and steepest group/cluster slopes reported in the literature 
\citep{Pisano2011,Martin2011,Stierwalt2009,Zwaan2005,RosenbergSchneider2002}. 

4. We do not find a statistically significant difference in the WF distribution of 
wall and void galaxies. These distributions are not well fit by a modified Schechter function. 

5. We find that the WF varies with galaxy colour with bluer galaxies increasing the low velocity slope. 
The WF of the blue and red galaxies are also not well fit by a modified Schechter function.

\section*{Acknowledgements} 

The authors would like to acknowledge the work of the entire
ALFALFA collaboration team in observing, flagging, and extracting 
the catalog of galaxies used in this work. 
We thank the referee for helpful comments and suggestions. 
The ALFALFA team at Cornell is supported by NSF grants AST-0607007 
and AST-1107390 to RG and MPH and by grants from the Brinson Foundation.

Funding for the creation and distribution of the SDSS Archive has been
provided by the Alfred P. Sloan Foundation, the Participating
Institutions, the National Aeronautics and Space Administration, the
National Science Foundation, the U.S. Department of Energy, the
Japanese Monbukagakusho, and the Max Planck Society. The SDSS Web site
is http://www.sdss.org/.

The SDSS is managed by the Astrophysical Research Consortium (ARC) for
the Participating Institutions. The Participating Institutions are The
University of Chicago, Fermilab, the Institute for Advanced Study, the
Japan Participation Group, The Johns Hopkins University, the Korean
Scientist Group, Los Alamos National Laboratory, the
Max-Planck-Institute for Astronomy (MPIA), the Max-Planck-Institute
for Astrophysics (MPA), New Mexico State University, University of
Pittsburgh, Princeton University, the United States Naval Observatory,
and the University of Washington.

\bibliographystyle{STYFiles/mn2e_fixed}
\bibliography{bibliography1}

\end{document}